\documentclass[aps,prd,preprint,superscriptaddress]{revtex4-2}
\usepackage{amsmath}
\usepackage{hyperref}
\usepackage{graphicx}
\usepackage{amsfonts}

\begin{document}


\title{Localized chaos due to rotating shock waves in Kerr-AdS black holes and their ultraspinning version}


\author{Hadyan Luthfan Prihadi}

\email{hadyan.luthfan@itb.ac.id}
\affiliation{Theoretical High Energy Physics Group, Department of Physics, FMIPA, Institut Teknologi Bandung, Jl. Ganesha 10 Bandung, Indonesia.}
\affiliation{Indonesia Center for Theoretical and Mathematical Physics (ICTMP), Institut Teknologi Bandung, Jl. Ganesha 10 Bandung,
	40132, Indonesia.}
	
\author{Freddy Permana Zen}
\email{fpzen@fi.itb.ac.id}
\affiliation{Theoretical High Energy Physics Group, Department of Physics, FMIPA, Institut Teknologi Bandung, Jl. Ganesha 10 Bandung, Indonesia.}
\affiliation{Indonesia Center for Theoretical and Mathematical Physics (ICTMP), Institut Teknologi Bandung, Jl. Ganesha 10 Bandung,
	40132, Indonesia.}
\author{Donny Dwiputra}
\email{donny.dwiputra@ymail.com}
\affiliation{Asia Pacific Center for Theoretical Physics, Pohang University of Science and Technology, Pohang 37673, Gyeongsangbuk-do, South Korea.}
\author{Seramika Ariwahjoedi}
\email{sera001@brin.go.id}
\affiliation{Asia Pacific Center for Theoretical Physics, Pohang University of Science and Technology, Pohang 37673, Gyeongsangbuk-do, South Korea.}
\affiliation{Research Center for Quantum Physics, National Research and Innovation Agency (BRIN), South Tangerang 15314, Indonesia.}
\date{\today}
\begin{abstract}
The butterfly velocity of four-dimensional rotating charged asymptotically AdS black hole is calculated to probe chaos using localized rotating shock waves. In this work, we obtain the angular momentum dependence of the butterfly velocity due to rotation in the shock wave probes. In general, the angular momentum $\mathcal{L}$ of the shock waves increases the butterfly velocity. The localized shocks also generate butterfly velocities which vanish when we approach extremality, indicating no entanglement spread near extremality. One of the butterfly velocity modes is well bounded by both the speed of light and the Schwarzschild-AdS result, while the other may become superluminal. Aside from the logarithmic behavior of the scrambling time which indicates chaos, the Lyapunov exponent is also positive and bounded by $\kappa=2\pi T_H/(1-\mu\mathcal{L})$. The Kerr-NUT-AdS and Kerr-Sen-AdS solutions and their ultraspinning versions are used as examples to attain a better understanding of the chaotic phenomena in rotating black holes, especially those with extra conserved charges. 
\end{abstract}


\maketitle

\section{Introduction}
The study of chaotic behavior of some quantum field theories (especially conformal field theories/CFT) which are conjectured to be dual to black holes has been an interesting and important topic since Shenker and Stanford \cite{Shenker2014} calculated the scrambling time and the Lyapunov exponent of a BTZ black hole. The Lyapunov exponent of a chaotic system was then found to be bounded from above by its temperature, or surface gravity in the black hole language \cite{Maldacena2016bound}. Saturation of the upper bound is then suggested to be the sufficient condition of a theory with an Einstein gravity dual. The holographic description \cite{Maldacena1997_LargeN,RyuTakayanagi1,RyuTakayanagi2,Hubeny_2007,Hartman2013} of chaotic behavior of various black hole spacetimes have been extensively studied \cite{Shenker2014, Leichenauer2014,Roberts2015, Jahnke2018,Avila2018,Fischler2018,Ahn2019,Jahnke2019,Poojary2020,Banerjee2020,Blake2022,Malvimat2022,Malvimat2023KerrAdS4,Malvimat2023,Amano2023, Prihadi2023} until very recently. Other than using the holographic principle, chaos in black holes can also be studied using other methods such as the geodesic analysis \cite{Zhao2018,Gwak2022,Yu2022,Yu2023,HE2023101325,Chen2023}. This work extends the holographic description of chaos in more general rotating and charged Kerr-AdS black holes in four dimensions due to rotating localized perturbations.\\
\indent An eternal black hole in Anti-de Sitter (AdS) background is suggested to have a CFT dual description described by an entangled thermofield double (TFD) state \cite{MaldacenaEternal2003}. This is an alternative description to the near-horizon microscopic description of a black hole via the Kerr/CFT correspondence \cite{Guica2009,Sakti2018,Sakti2019,Sakti2020b,Sakti2020,Sakti2021,Sakti2022}. The chaotic behavior of a black hole is studied by disrupting the entanglement pattern of the TFD state by an operator in the far past. In the gravity description, the operator is viewed as a gravitational shock wave sent from the past traveling very close to the speed of light near the horizon. Although the shock wave initially has a $\mathcal{O}(\hbar)$ energy, it is then highly blueshifted and becomes of order one before entering the black hole. This process can be described by the shock wave geometry or the Dray-'t Hooft solution. The signature of chaos appears when the four-point out-of-time-ordered correlator (OTOC) approaches zero exponentially and vanishes at time scale $\tau_*$, which is defined to be the scrambling time \cite{Shenker2014}. For a system with scrambling time proportional to the logarithmic of the system's entropy, the system is said to be a fast scrambler, and black holes are conjectured to be the fastest scrambler in nature \cite{Sekino2008}. In the holographic description, the OTOC is calculated using the holographic mutual information with the Ryu-Takayanagi \cite{RyuTakayanagi1,RyuTakayanagi2} and the Hartman-Maldacena \cite{Hartman2013} surfaces. The OTOC is related to the shock wave profile that enters the black hole \cite{Shenker2014,Roberts2015}.\\
\indent By perturbing the TFD state using local operators, we can also extract information about the effective "light cone" in which the OTOC can only vanish inside it \cite{Roberts2015}. Such a speed of propagation is defined as the butterfly velocity $v_B$. For a static black hole in AdS, the butterfly velocity solely depends on the spacetime dimension in the large black hole limit \cite{Shenker2014,Roberts2015}.\\
\indent More general black holes in AdS should also indicate chaos in their boundary theories, including the one with rotation and charges. The uncharged cases were first studied using holography for the Kerr-AdS black hole in 4 and 5 spacetime dimensions \cite{Malvimat2023KerrAdS4,Malvimat2023} and has recently been extended to the Kerr-Sen-AdS$_4$ black hole as the solution to the Einstein-Maxwell dilaton-axion (EMDA) theory with electromagnetic, dilaton, and axion charges \cite{Prihadi2023}. However, the properties of chaos for more general Kerr-AdS black holes due to rotating and charged shock waves are still incomplete, especially important information about how local entanglement disruption spread throughout the spacetime is still lacking. The use of rotating (and charged) shock waves might also give us interesting result since it is known that rotating shock waves give a new bound for the Lyapunov exponent \cite{Malvimat2022,Malvimat2023KerrAdS4,Prihadi2023}. We also consider the ultraspinning limit $a\rightarrow l$, where $a$ is the rotation parameter of the black hole and $l$ is the AdS radius, of the Kerr-AdS black hole solutions. This limit is unique to the rotating AdS black holes and important since the solution will be different compared to the standard Kerr-AdS black hole and not just simply obtained by taking the $a\rightarrow l$ limit of the standard Kerr-AdS black hole solutions.\\
\indent The study of the chaotic properties of more general rotating black hole is crucial to further understand the dual CFT theory and gives us insights into understanding the quantum aspects of gravity via holography, which is inspired by string theory. Other example of previous study about quantum gravity is loop quantum gravity \cite{Rovelli1990,Ariwahjoedi2015}.\\
\indent In earlier works, the rotating shock waves is first used as a probe to calculate the Lyapunov exponent of a rotating BTZ black hole \cite{Malvimat2022} and rotating Kerr-AdS black holes in 4 and 5 dimensions \cite{Malvimat2023KerrAdS4, Malvimat2023}. Later on, the rotating and charged shock waves is used to calculate the Lyapunov exponent and scrambling time delay of a Kerr-Sen-AdS black hole \cite{Prihadi2023}. \\
\indent The main distinction between our work and previous works about local chaos in black holes (such as \cite{Jahnke2018,Mezei2020}) is that our result depends on the angular momentum of the rotating shock waves $\mathcal{L}$ while previous analysis only use non-rotating shock waves. We find that the butterfly velocity of the rotating black hole in AdS is highly influenced by the black hole angular momentum $\mu$ and the shock waves angular momentum $\mathcal{L}$, as well as the black hole charges. We use the Kerr-NUT-AdS and Kerr-Sen-AdS solutions as examples to test the behavior of the butterfly velocity. Our result indicates that the angular velocity of the rotating shock waves $\mathcal{L}$ increases the butterfly velocity of the black hole, similar to the Lyapunov exponent that is also increased due to the presence of $\mathcal{L}$ \cite{Malvimat2022, Malvimat2023KerrAdS4, Malvimat2023, Prihadi2023}.\\
\indent The content structure of this work is as follows: In Sec. \ref{sec2}, we briefly write the solutions to the Kerr-NUT-AdS and the Kerr-Sen-AdS black holes that serve as the background in calculating the butterfly velocity. In Sec. \ref{sec3}, we derive the Kruskal coordinates from the rotating and charged shock waves perspective. In Sec. \ref{sec4}, we calculate the butterfly velocity of the Kerr-AdS black holes from solving the Einstein's equation sourced by both matter and the localized shock waves. This is the main result of this work. In Sec. \ref{sec5}, we calculate the butterfly velocities for the ultraspinning cases. We then plot the Lyapunov exponent of both Kerr-NUT-AdS and Kerr-Sen-AdS black holes to make comparison between the two in Sec. \ref{sec6}. We summarize our results in the Conclusions and Discussions session in Sec. \ref{sec7}.
\section{The Kerr-AdS Black Holes and their Ultraspinning Limit}\label{sec2}
In this section, we briefly review the metric of the Kerr-AdS black holes that are used in this work as the background. The first black hole solution is the charged Kerr-NUT-AdS solution \footnote{The electrically and magnetically charged Kerr-NUT-AdS black hole is called the Kerr-Newman-NUT-AdS black hole. However, in this work, we refer the Kerr-NUT-AdS to the charged version as well for brevity.}. This solution can be obtained from the general Petrov type D solution of the Einstein's equation with vanishing acceleration parameter \cite{Griffiths2006}. The metric in Boyer-Lindquist coordinates is given by \cite{Rodriguez2022}
\begin{equation}
ds^2=-\frac{\Delta}{\Sigma}X^2+\frac{\Sigma}{\Delta}dr^2+\frac{\Sigma}{\Delta_\theta}d\theta^2+\frac{\Delta_\theta\sin^2\theta}{\Sigma}Y^2,
\end{equation}
where
\begin{align}
X&=dt+\frac{(2n\cos\theta-a\sin^2\theta)}{\Xi} d\varphi,\\\nonumber
Y&=adt-\frac{(r^2+n^2+a^2)}{\Xi}d\varphi,
\end{align}
\begin{align}
\Delta(r)&=\frac{3(a^2-n^2)n^2+(a^2+6n^2)r^2+r^4}{l^2}\\\nonumber&\;\;\;\;\;\;+r^2+a^2-2mr-n^2+p^2+q^2,
\end{align}
\begin{align}
\Delta_\theta&=1-\frac{4an\cos\theta}{l^2}-\frac{a^2\cos^2}{l^2},\;\;\;\Xi=1-\frac{a^2}{l^2},\\\nonumber
\Sigma&=r^2+(n+a\cos\theta)^2.
\end{align}
The function $\Delta(r)$ determines the location of the horizon, with the outermost horizon $r_+$ is the largest solution to $\Delta(r)=0$. The parameters in the Kerr-NUT-AdS black holes are $\{m,a,p,q,n,l\}$ which describe its mass, rotation, magnetic charge, electric charge, NUT parameter, and the AdS radius respectively. \\
\indent On the other hand, the Kerr-Sen-AdS black hole solution is obtained from the equation of motion derived from the gauged Einstein-Maxwell dilaton-axion action \cite{Wu2021}. The solution to the equation of motion is given by
\begin{equation}
ds^2=-\frac{\bar{\Delta}}{\bar{\Sigma}}\bar{X}^2+\frac{\bar{\Sigma}}{\bar{\Delta}}dr^2+\frac{\bar{\Sigma}}{\bar{\Delta}_\theta}d\theta^2+\frac{\bar{\Delta}_\theta\sin^2\theta}{\bar{\Sigma}}\bar{Y}^2,
\end{equation}
where
\begin{align}
\bar{X}&=dt-\frac{d\sin^2\theta}{\Xi} d\varphi,\\\nonumber
\bar{Y}&=adt-\frac{(r^2-d^2-k^2+a^2)}{\Xi}d\varphi,
\end{align}
\begin{align}
\bar{\Delta}(r)=&\bigg(1+\frac{r^2-d^2-k^2}{l^2}\bigg)(r^2-d^2-k^2+a^2)\\\nonumber
&-2mr+p^2+q^2
\end{align}
\begin{align}
\bar{\Delta}_\theta&=1-\frac{a^2\cos^2}{l^2},\;\;\;\Xi=1-\frac{a^2}{l^2},\\\nonumber
\bar{\Sigma}&=r^2-d^2-k^2+a^2\cos^2\theta.
\end{align}
Again, the function $\bar{\Delta}(r)$ determines the horizon and the Kerr-Sen-AdS black hole is characterized by $\{m,a,p,q,d,k,l\}$ which describe its mass, rotation, magnetic charge, electric charge, dilaton charge, axion charge, and AdS radius respectively. \\
\indent The thermodynamics quantities of the Kerr-NUT-AdS and the Kerr-Sen-AdS black hole are given by
\begin{align}
M&=\frac{m}{\Xi},\;\;\;J=\frac{ma}{\Xi},\;\;\;Q=\frac{q}{\Xi},\;\;\;P=\frac{p}{\Xi},\\\nonumber
S&=\frac{\pi}{\Xi}(r_+^2+\Upsilon^2+a^2),\\\nonumber
\Omega_\varphi&=\frac{a\Xi}{r_+^2+\Upsilon^2+a^2},\\\nonumber
T_H&=\frac{\Delta'(r_+)}{r_+^2+\Upsilon^2+a^2},
\end{align}
where the function $\Upsilon^2$ is defined to make some distinction between the Kerr-NUT-AdS and the Kerr-Sen-AdS black holes and it is given by
\begin{equation}
    \Upsilon^2= 
\begin{cases}
    n^2,& \text{Kerr-NUT-AdS}\\
    -d^2-k^2,&\text{Kerr-Sen-AdS}\\
    0,              & \text{Kerr-AdS}
\end{cases}
\end{equation}
The thermodynamics functions denotes the black hole's mass $M$, conserved angular momentum $J$, conserved electric charge $Q$, and conserved magnetic charge $P$. The entropy of the black hole is given by $S$ while $\Omega_\varphi$ and $T_H$ are the horizon's angular velocity and the Hawking temperature respectively.\\
\indent The ultraspinning black hole solutions can be obtained by redefining the coordinate $\varphi\rightarrow\varphi/\Xi$ and take the $a\rightarrow l$ limit. The important difference between the standard Kerr-AdS black holes and the ultraspinning versions that is used in this work is the horizon's angular velocity. It is now given by
\begin{equation}
\hat{\Omega}_\varphi=\frac{l}{\hat{r}_+^2+\Upsilon^2+l^2},
\end{equation}
which works for both the Kerr-NUT-AdS and the Kerr-Sen-AdS black holes. The outer horizon $\hat{r}_+$ is the largest solution to the $a\rightarrow l$ limit of either $\Delta(r)=0$ or $\bar{\Delta}(r)=0$, depending on which background is being used. Therefore, $\hat{r}_+$ varies with the type of the black hole.
\section{Kruskal Coordinates from Rotating Charged Shock Waves}\label{sec3} 
Perturbation of the entangled CFT dual of a rotating and charged eternal Kerr-AdS black hole corresponds to gravitational shock waves traveling from the boundary to the interior of the black hole. The geometry of the traveling shock waves is represented by the Dray-'t Hooft solution \cite{Dray1985,Dray1985b}. The shock waves geometry due to rotating shock waves with angular momentum $\mathcal{L}$ was first studied in \cite{Malvimat2022} for a three-dimensional BTZ black hole and recently extended to a four-dimensional Kerr-AdS \cite{Malvimat2023KerrAdS4} and a Kerr-Sen-AdS black hole \cite{Prihadi2023}. In this work, such a shock wave geometry is used to probe the localized entanglement disruption of Kerr-AdS black hole families. The gravitational shock waves follow an approximately null-like path denoted by $\xi_\pm$.
\subsection{Metric in the Kruskal Coordinates}
\indent The geodesics $\xi_+$ of a null rotating shock wave with energy $\mathcal{E}$ and angular momentum $\mathcal{L}$ satisfy \cite{Malvimat2023KerrAdS4}
\begin{equation}
\xi_+^2=0,\;\;\;\xi_+\cdot\zeta_t=-\mathcal{E},\;\;\;\xi_+\cdot\zeta_\varphi=\mathcal{L},\;\;\;\xi_+^\mu K_{\mu\nu}\xi_+^\nu=Q,
\end{equation}
where $\zeta_t=\partial_t-a/l^2\partial_\varphi$ and $\zeta_\varphi=\partial_\varphi$ are the Killing vectors correspond to time translation and rotational invariant respectively and $K_{\mu\nu}$ is the Killing-Yano tensor with Carter's constant $Q$. The explicit form of $K_{\mu\nu}$ and $Q$ is not important and we can solve the geodesics by imposing the boundary condition $\xi^\theta=0$ at the equator $\theta=\pi/2$. From axisymmetry, we can have other geodesic solutions denoted as $\xi_-$ from reversing the sign of $-\mathcal{E}\rightarrow\mathcal{E}$ and $\mathcal{L}\rightarrow-\mathcal{L}$.\\
\indent The geodesics are then used to write the Kruskal-like coordinates for equatorial rotating shock waves in the Kerr-AdS background, which is given by
\begin{align}
\xi_+\cdot dx&=dr_*-d\tau-\xi_\theta d\theta\equiv du,\\
\xi_-\cdot dx&=dr_*+d\tau+\xi_\theta d\theta\equiv dv,
\end{align} 
where $\tau=(1-a\mathcal{L}/l^2)t-\mathcal{L}\varphi$. The tortoise-like coordinate $r_*$ is given by 
\begin{equation}
r_*(r)=\int\frac{\tilde{f}}{\Delta}dr,
\end{equation}
where
\begin{equation}
\tilde{f}^2=-\Delta(\mathcal{L}-a)^2+(\mathcal{L}a(1+(r^2+\Upsilon^2)/l^2)-(r^2+\Upsilon^2+a^2))^2,
\end{equation}
and the theta component of the geodesics is given by
\begin{equation}
\xi_\theta=\frac{\sqrt{-\Xi\cot^2\theta(l^2(\Xi-\Delta_\theta)+\mathcal{L}^2\Delta_\theta)}}{\Delta_\theta}.
\end{equation}
The metric of the Kerr-AdS black hole in this Kruskal-like coordinates can then be written as
\begin{align}
ds^2=&F(r,\theta)dudv+h(r,\theta)(dz+h_\tau(r,\theta)d\tau)^2\\&+g(r,\theta)(d\theta+g_\tau(r,\theta)d\tau)^2,
\end{align}
where the explicit form of the functions $\{F,h,h_\tau,g,g_\tau\}$ are given by
\begin{widetext}
\begin{align}
F(r,\theta)=&\frac{\Delta\Sigma}{\tilde{f}^2},\\\nonumber
h(r,\theta)=&\eta^2\frac{(\Delta_\theta\sin^2\theta(\mathcal{L}a(1+(r^2+\Upsilon^2)/l^2)-(r^2+\Upsilon^2+a^2))^2-\Delta(\mathcal{L}\Delta_\theta-a\sin^2\theta)^2)}{\Sigma(1-a\mathcal{L}/l^2)^2\Xi^2},\\\nonumber
h_\tau(r,\theta)=&\eta^{-1}\bigg[\frac{\Xi(a\sin^2\theta\Delta_\theta(a\mathcal{L}(1+(r^2+\Upsilon^2)/l^2)-(r^2+\Upsilon^2+a^2))-\Delta(\mathcal{L}\Delta_\theta-a\sin^2\theta))}{\Delta_\theta\sin^2\theta(\mathcal{L}a(1+(r^2+\Upsilon^2)/l^2)-(r^2+\Upsilon^2+a^2))^2-\Delta(\mathcal{L}\Delta_\theta-a\sin^2\theta)^2}+\gamma\bigg],\\\nonumber
g(r,\theta)=&\frac{\Sigma(\Delta_\theta\sin^2\theta(\mathcal{L}a(1+(r^2+\Upsilon^2)/l^2)-(r^2+\Upsilon^2+a^2))^2-\Delta(\mathcal{L}\Delta_\theta-a\sin^2\theta)^2)}{\tilde{f}^2\Delta_\theta^2\sin^2\theta},\\\nonumber
g_\tau(t,\theta)=&\frac{\Delta\Delta_\theta\sin^2\theta\sqrt{-\Xi\cot^2\theta(l^2(\Xi-\Delta_\theta)+\Delta_\theta\mathcal{L}^2)}}{\Delta_\theta\sin^2\theta(\mathcal{L}a(1+(r^2+\Upsilon^2)/l^2)-(r^2+\Upsilon^2+a^2))^2-\Delta(\mathcal{L}\Delta_\theta-a\sin^2\theta)^2}.
\end{align}
\end{widetext}
\indent Here, we have performed a coordinate transformation $\varphi\rightarrow\eta z+\gamma\tau$ such that $h_\tau(r)$ behaves as $\mathcal{O}(r-r_+)$ near the horizon and recover the black hole horizon area when $z$ is integrated from 0 to $2\pi$, following \cite{Malvimat2023KerrAdS4,Prihadi2023}. In this case, we have $\eta=\frac{1}{1-\mu\mathcal{L}}$ and $\gamma=\frac{\Omega_\varphi}{1-\mu\mathcal{L}}$ with $\mu\equiv\Omega_\varphi+a/l^2$.\\
\indent The metric in the affine coordinates at $r_+$, $U=-e^{\kappa u}$ and $V=e^{\kappa v}$, where
$
\kappa=\frac{2\pi T_H}{1-\mu\mathcal{L}},
$
can be written as
\begin{align}\label{metricUV}
ds^2=&\frac{F}{\kappa^2UV}dUdV\\\nonumber
&+h\bigg(dz+\frac{h_\tau}{2\kappa UV}(UdV-VdU)-h_\tau\xi_\theta d\theta\bigg)^2\\\nonumber
&+g\bigg(d\theta+\frac{g_\tau}{2\kappa UV}(UdV-VdU)-g_\tau\xi_\theta d\theta\bigg)^2.
\end{align}
We use this coordinate to generate the Dray-'t Hooft solution due to the rotating and charged shock waves. Later on, we write the functions as $\{F(r),h(r),h_\tau(r),g(r),g_\tau(r)\}$ for the equatorial case with $\theta=\pi/2$.
\subsection{Location of the Asymptotic}
Consider again the Kerr-AdS black hole metric in Kruskal coordinates at the equator with $\theta=\pi/2$, which is given by
\begin{equation}\
ds^2=\frac{F}{\kappa^2 UV}dUdV+h(dz+h_\tau d\tau)^2.
\end{equation}
In this subsection, we would like to derive the asymptotic metric with the existence of the shock waves angular momentum $\mathcal{L}$. At the asymptotic, the radial value becomes $r\rightarrow r_c\gg1$. The functions in the metric then becomes
\begin{equation}
\tilde{f}(r_c)^2\approx r_c^4\Xi(1-\mathcal{L}^2/l^2),
\end{equation}
\begin{equation}
h(r_c)\approx\frac{r_c^2(1-\mathcal{L}^2/l^2)}{\Xi(1-\mu\mathcal{L})^2(1-a\mathcal{L}/l^2)^2},
\end{equation}
\begin{equation}
h_\tau(r_c)\approx(1-\mu\mathcal{L})\bigg[\frac{(a-\mathcal{L})/l^2}{1-\mathcal{L}/l^2}+\frac{\Omega_\varphi}{1-\mu\mathcal{L}}\bigg],
\end{equation}
\begin{equation}
F(r_c)\approx\frac{r_c^2/l^2}{\Xi(1-\mathcal{L}/l^2)}.
\end{equation}
From this asymptotic limit, the Kerr-AdS black hole metric becomes
\begin{align}\label{asympmetric}
ds^2\approx&\frac{r_c^2/l^2}{\Xi(1-\mathcal{L}/l^2)}\bigg[-d\tau^2\\\nonumber
&+l^2\bigg(\frac{(1-\mathcal{L}^2/l^2)}{(1-\mu\mathcal{L})(1-a\mathcal{L}/l^2)}\bigg)^2(dz-\omega d\tau)^2\bigg].
\end{align}
The metric inside the square bracket denotes flat space-time in a polar coordinates with radius
\begin{equation}\label{radasymptotics}
\bar{r}_c(\mathcal{L})=l\bigg(\frac{(1-\mathcal{L}^2/l^2)}{(1-\mu\mathcal{L})(1-a\mathcal{L}/l^2)}\bigg),
\end{equation}
and rotating with angular velocity $\omega=h_\tau(r_c)$. The radius of the asymptotic determines $\bar{r}_c(\mathcal{L})$ as the prefactor of the butterfly velocity which will be represented later on in order to represent the velocity as a tangential velocity at the asymptotic instead of an angular velocity. The radius $\bar{r}_c(\mathcal{L})$ depends on the angular velocity of the shock waves $\mathcal{L}$ and reduce to the standard radius in black hole with AdS background $\bar{r}_c(0)=l$ when the angular momentum vanishes.
\section{Localized Dray-'t Hooft Solution and Butterfly Velocities}\label{sec4} 
\subsection{Einstein's Equation}
\indent The shock wave solution can be obtained by using the shift in the $V$ coordinate such that $V\rightarrow V+\alpha f(z,\theta)\Theta(U-U_0)$, where $\Theta(U-U_0)$ denotes the Heaviside step function. The value of the shift is expressed by the parameter $\alpha f(z,\theta)$, which can be a function of $\tau$, $\theta$ and $z$. If we choose the perturbations to  always stay at the equator, then $\alpha f(z)$ can only be the function of $\tau$ and $z$. After the coordinate shift, the metric becomes
\begin{equation}\label{metricshift}
ds^2\rightarrow \tilde{ds}^2-\bigg(\frac{F}{\kappa^2 UV}\bigg)\alpha f(z)\delta(U)dU^2.
\end{equation}
This solves the Einstein equation with cosmological constant $\Lambda=-\frac{3}{l^2}$ sourced by both matter and shock waves energy-momentum tensor
\begin{equation}
R_{\mu\nu}-\frac{1}{2}R g_{\mu\nu}+\Lambda g_{\mu\nu}=8\pi G_N(T^{\text{matter}}_{\mu\nu}+T^{\text{shocks}}_{\mu\nu}).\label{Einstein}
\end{equation}
The form of $T_{\mu\nu}^{\text{shocks}}$ depends on the shock waves profile. For localized equatorial shock waves along the $U=const.$ trajectory, the energy-momentum of the shocks can be written as
\begin{equation}\label{shocksT}
T_{UU}^{\text{shocks}}=\mathbb{B}\alpha\delta(z-z_0)\delta(U)dU^2,
\end{equation}
where $\mathbb{B}$ is some constant that depends on the energy of the shock waves and we choose point source $\delta(z-z_0)$ to describe the localized shock wave profile. Here we assume that the shock waves always stay along the equator and hence its profile does not depend on $\theta$.\\
\indent The energy-momentum tensor for the matter fields is given by
\begin{align}
T^{\text{matter}}_{\mu\nu}dx^\mu &dx^\nu\\\nonumber=&T_{UU}dU^2+2T_{UV}dUdV+T_{VV}dV^2\\\nonumber&+T_{zz}dz^2+2T_{Uz}dUdz+2T_{Vz}dVdz.
\end{align}
For rotating black hole with axial symmetry, there are crossing terms between time coordinate $t$ and the axial coordinate $\varphi$. The crossing term is depicted by both $Vz$ and $Uz$ components of the energy-momentum tensor. After the shift in the $V$-direction, the energy-momentum tensor becomes
\begin{align}
T&^{\text{matter}}_{\mu\nu}dx^\mu dx^\nu\\\nonumber=&(T_{UU}-2T_{UV}\alpha f(z)\delta(U)+T_{VV}\alpha^2f(z)^2\delta(U)^2)dU^2\\\nonumber&+(2T_{UV}-2T_{VV}\alpha f(z)\delta(U))dUdV\\\nonumber&(2T_{Uz}-2T_{Vz}\alpha f(z)\delta(U))dUdz\\\nonumber&+T_{zz}dz^2+2T_{Vz}dVdz.
\end{align}
\\
\indent The value of $\alpha$ can be obtained by considering the smoothness of the geometry at the horizon and using the first-law of thermodynamics for the rotating AdS black hole. From the calculations performed in \cite{Prihadi2023}, we have
\begin{equation}
\alpha=\varepsilon e^{\kappa\tau_0},
\end{equation}
where $\varepsilon$ depends on the energy of the shock waves. Upon taking the double-scaling limit $\varepsilon\rightarrow0$ and $e^{\kappa\tau_0}\rightarrow\infty$, the value of $\alpha$ can be of order 1 and becomes significant at late times. But first, in deriving the Einstein's equation, we treat the function $\alpha$ perturbatively.\\
\indent In order to obtain the function $f(z)$, we need to solve the Einstein's field equation sourced by both $T^{\text{matter}}$ and $T^{\text{shocks}}$ in eq. (\ref{Einstein}). In order to do so, we treat the parameter $\alpha$ as a small parameter by scaling $\alpha\rightarrow \varepsilon\alpha$ and $T^{\text{shocks}}\rightarrow\varepsilon T^{\text{shocks}}$ in which $\varepsilon=0$ reproduces the unperturbed equation of motion. The Einstein's equation sourced only by $T^{\text{matter}}$ is automatically satisfied at $\varepsilon=0$ and we take the solution to be the Kerr-AdS black hole families. \\
\indent We then expand the Einstein's equation in the power of $\varepsilon$. The $UV$ component of the Einstein's equation at $\mathcal{O}(\varepsilon)$ and the $UU$ component at $\mathcal{O}(\varepsilon^2)$ are respectively given by
\begin{align}
\bigg[\bigg(&\frac{F}{\kappa^2 UV}\bigg)_{,V}\bigg(\frac{\kappa^2 UV}{F}\bigg)\frac{h_{,V}}{2h}+\frac{h_{,V}^2}{2h^2}-\frac{h_{,VV}}{h}\bigg]\delta(U)f(z)\alpha\\\nonumber
&=8\pi G_N T_{VV}\delta(U)f(z)\alpha,
\end{align}
and
\begin{align}
\bigg[\bigg(\frac{F}{\kappa^2 UV}\bigg)_{,V}&\bigg(\frac{\kappa^2 UV}{F}\bigg)\frac{h_{,V}}{h}+\frac{h_{,V}^2}{h^2}-\frac{2h_{,VV}}{h}\bigg]\delta(U)^2f(z)^2\alpha^2\\\nonumber
=&8\pi G_N T_{VV}\delta(U)^2f(z)^2\alpha^2.
\end{align}
From these two equations, we set \cite{Jahnke2018}
\begin{equation}
\bigg(\frac{F}{\kappa^2UV}\bigg)_{,V}=h_{,V}=h_{,VV}=0,
\end{equation}
at the horizon and hence $T_{VV}=0$ at the horizon. Using those conditions, the $Uz$ component at $\mathcal{O}(\varepsilon)$ and the $Vz$ component at $\mathcal{O}(1)$ of the Einstein's equation are respectively given by
\begin{align}\label{Uz1}
\bigg[&\bigg(\frac{\kappa^2UV}{F}\bigg)\bigg(\frac{h_\tau}{2\kappa UV}\bigg)_{,V}(6h)-\bigg(\frac{\kappa^2 UV}{F}\bigg)^2\bigg(\frac{h_\tau}{2\kappa UV}\bigg)\\\nonumber
&\bigg(\frac{F}{\kappa^2 UV}\bigg)_{,VV}(2hV)+\bigg(\frac{\kappa^2 UV}{F}\bigg)\bigg(\frac{h_\tau}{2\kappa UV}\bigg)_{,VV}(2hV)\bigg]\\\nonumber
&\;\;\;\times\delta(U)f(z)\alpha=-8\pi G_N T_{Vz}\delta(U)f(z)\alpha,
\end{align}
and
\begin{align}
-&\bigg(\frac{\kappa^2 UV}{F}\bigg)\bigg(\frac{h_\tau}{2\kappa UV}\bigg)_{,V}(3h)-\bigg(\frac{\kappa^2UV}{F}\bigg)\bigg(\frac{h_\tau}{2\kappa UV}\bigg)_{,VV}\\\nonumber
&\times(hV)=8\pi G_NT_{Vz}.
\end{align}
The $Vz$ component with $\mathcal{O}(1)$ is automatically satisfied by the equation of motion. Thus, we can insert the value of $T_{Vz}$ into eq. (\ref{Uz1}). This gives us
\begin{equation}
\bigg(\frac{h_\tau}{2\kappa UV}\bigg)_{,V}=\bigg(\frac{h_\tau}{2\kappa UV}\bigg)_{,VV}=\bigg(\frac{F}{\kappa^2 UV}\bigg)_{,VV}=0,
\end{equation}
at the horizon, and hence we also have $T_{Vz}=0$ at the horizon. The other components of the Einstein's equation with $\mathcal{O}(1)$ are automatically satisfied by the equation of motion. Another important equation at $\mathcal{O}(1)$ is the $UV$ component at $\mathcal{O}(1)$, which is given by
\begin{align}
-\frac{1}{2l^2}&\bigg(\frac{F}{\kappa^2 UV}\bigg)+\bigg(\frac{\kappa^2 UV}{F}\bigg)\bigg(\frac{h_\tau}{2\kappa UV}\bigg)^2(2h)+\frac{h_{,UV}}{2h}\\\nonumber
&=8\pi G_N T_{UV}.
\end{align}
Here, we obtain the value of $T_{UV}$, which can be inserted into the $UU$ component at $\mathcal{O}(\varepsilon)$. By inserting the $T_{UU}^{\text{shocks}}$ from eq. (\ref{shocksT}) into the $UU$ component of the Einstein's equation at $\mathcal{O}(\varepsilon)$, we have
\begin{align}\label{eom}
\bigg(&\frac{F}{\kappa^2UV}\bigg)\frac{1}{2h}f''(z)\delta(U)-\bigg(\frac{h_\tau}{\kappa UV}\bigg)f'(z)\delta(U)\\\nonumber
&-\frac{h_{,UV}}{h}f(z)\delta(U)=8\pi G_N\mathbb{B}\delta(z-z_0)\delta(U).
\end{align}
\indent All of the components in the equation are evaluated at the horizon since it is multiplied by $\delta(U)$. The relation between the coordinate $r$ and $UV$ is given by \cite{Prihadi2023}
\begin{equation}
(r-r_+)F'(r_+)=-\mathbb{A}UV,
\end{equation}
for some dimensionless constant $\mathbb{A}$. Therefore, we have
\begin{equation}
\bigg(\frac{F}{\kappa^2 UV}\bigg)_{U=0}\frac{1}{2h}\bigg|_{U=0}=-\frac{\mathbb{A}}{\kappa^2}\frac{1}{2h(r_+)},
\end{equation}
\begin{equation}
\bigg(\frac{h_\tau}{\kappa UV}\bigg)_{U=0}=-\frac{\mathbb{A}}{F'(r_+)}\frac{h'_\tau(r_+)}{\kappa},
\end{equation}
\begin{equation}
\frac{h_{,UV}}{2h}\bigg|_{U=0}=-\frac{\mathbb{A}}{F'(r_+)}\frac{h'(r_+)}{2h(r_+)}.
\end{equation}
The constant $\mathbb{A}$ can be put into the right hand side of eq. (\ref{eom}). The equation of motion for $f(z)$ reduces to the well-known result such as the one found in \cite{Jahnke2018} if we take the non-rotating and uncharged case.
\subsection{Localized Shocks and Butterfly Velocities}
\begin{center}
\begin{figure}
\vspace{5mm}
\includegraphics[scale=0.7]{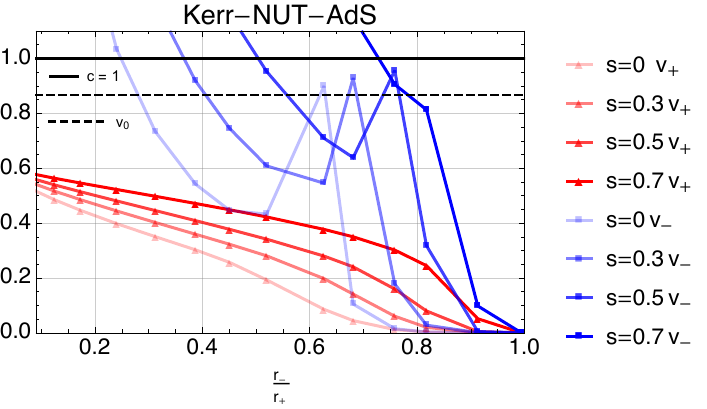}\\
\vspace{5mm}
\includegraphics[scale=0.7]{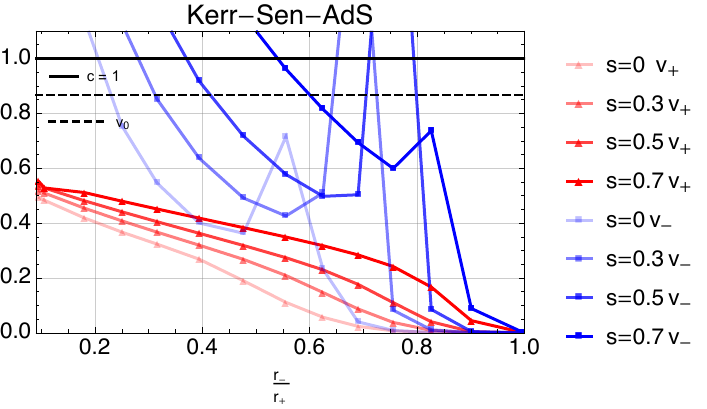}
\caption{\label{ButterflySen}Plot for the butterfly velocities $v_B^\pm$ for various angular momentum $\mathcal{L}$ which is scaled as $\mathcal{L}=s\mathfrak{r}/\mu$. We plot for $s=\{0,0.3,0.5,0.7\}$ from $s=0$ which is depicted by the line with lowest opacity up to $s=0.7$ which is the one with solid line. Blue lines represent $v_B^-$ while red lines represent $v_B^+$. The solid black line represent the speed of light $c=1$ while the dashed black line represent the Schwarzschild butterfly velocity bound $v_{B0}=\sqrt{3}/2$.}
\end{figure}
\end{center}

\indent From the previous derivations, the equation for $f(z)$ can be written as
\begin{equation}\label{eqfz}
(A\partial^2_z-B\partial_z-C)f(z)=\#\delta(z-z_0),
\end{equation}
where $\#$ denotes some constant that depends on the energy of the shock waves and
\begin{align}
A=\frac{1}{2\kappa^2h(r_+)},\;\;\;B=\frac{h'_\tau(r_+)}{\kappa F'(r_+)},\;\;\;C=\frac{h'(r_+)}{2h(r_+)}\frac{1}{F'(r_+)}.
\end{align}
Note that $z$ coordinate is periodic with periodicity $2\pi$. The solution to the differential equation of $f(z)$ reads
\begin{align}\label{fzresult}
\alpha f(z)\sim&\frac{2\pi}{w_+-w_-}\frac{e^{\kappa\tau_0-|w_+|[(z-z_0)(\text{mod }2\pi)]}}{1-e^{2\pi w_+}}\\\nonumber&+\frac{2\pi}{w_+-w_-}\frac{e^{\kappa\tau_0+|w_-|[(z-z_0)(\text{mod }2\pi)]}}{e^{2\pi w_-}-1},
\end{align}
up to some constant that depends on $\mathbb{B}$ and the other parameters of the black hole, with
\begin{equation}
w_\pm\equiv\frac{B}{2A}\pm\sqrt{\bigg(\frac{B}{2A}\bigg)^2+\frac{C}{A}}.
\end{equation}
Note that since $z$ is periodic with periodicity $2\pi$, the result found in eq. (\ref{fzresult}) only applies for $z\in [0,2\pi]$. Therefore, we should add the (mod $2\pi$) factor in eq. (\ref{fzresult}) since forgetting such a factor would lead to erroneous result in the growth of the OTOC \cite{Mezei2020}. The result shows us that the perturbations spread throughout the entire region with velocity (called the butterfly velocity) given by
\begin{equation}\label{butterflyvelocity}
v_B^{\pm}=\frac{l(1-\mathcal{L}^2/l^2)}{(1-\mu\mathcal{L})(1-a\mathcal{L}/l^2)}\frac{\kappa}{|w_\pm|}.
\end{equation}
The prefactor comes from the asymptotic radius of the Kerr-AdS metric in a rotating shock wave frame which is given by eq. (\ref{radasymptotics}). For the non-rotating, uncharged black hole and static shock wave limit $a,q,p,n,\mathcal{L}\rightarrow0$, the butterfly velocities reduce to the well-known universal result $v_{B0}=\sqrt{3}/2$ for a static black hole in 4 dimension \cite{Shenker2014}, in large black hole limit $r_+/l\gg1$.\\
\indent In contrast to the static case, for the rotating Kerr-AdS black hole, there is the $B$ term which will vanish in the static and neutral case. This term also causes the absolute value of positive $v_B^+$ and negative $v_B^-$ butterfly velocities to differ, thus breaking the symmetry of the butterfly velocities for static black holes. In the static case, the left- and right-moving butterfly velocities should only differ by a negative sign while in the rotating case, they are influenced by the direction of the rotation of the black hole and the shock waves. Such a term also appear in various calculations involving the rotating BTZ black hole \cite{Mezei2020,Malvimat2022}. For $\mathcal{L}=0$ and vanishing charges, the result agrees with the perturbative result found in \cite{Blake2022}, while in this work, we obtain the non-perturbative result of the butterfly velocity.\\
\indent For rotating cases, there are two kinds of butterfly velocities traveling in the opposite direction. Furthermore, the two velocities get shifted due to the rotation of the black holes such that $v_B^+\neq-v_B^-$. To see the behavior of $v_B^\pm$ for various black hole parameter values, we plot the result as a function of the ratio $\mathfrak{r}\equiv\frac{r_-}{r_+}$ in Fig. \ref{ButterflySen}. In this case, we scale the angular velocity of the shock waves as $\mathcal{L}=s\mathfrak{r}/\mu$ and vary $s$. We see that at some values of black hole parameters, the negative modes of the butterfly velocities $v_B^-$ become greater than the speed of light $c=1$ while the positive mode $v_B^+$ is well bounded by both $c$ and the Schwarzschild-AdS limit $v_{B0}=\sqrt{3}/2$. This $v_B^->c$ phenomena also appear in the BTZ case where one of the butterfly velocity becomes greater than the speed of light \cite{Jahnke2019}. \\
\indent In the extremal limit $\mathfrak{r}\rightarrow1$, the butterfly velocities quickly approach zero, indicating that chaos does not spread in the rotating extremal black holes in AdS. Furthermore, increasing the shock waves' angular momentum $\mathcal{L}$ opts to increase the butterfly velocity. The calculation of the butterfly velocity done in \cite{Blake2022} has some drawbacks since it cannot investigate how localized perturbation spreads in the extremal limit.\\
\indent Some other remarks can be made from the result of the butterfly velocities. Both Kerr-Sen-AdS and Kerr-NUT-AdS give us similar behaviors. In the $\mathfrak{r}\rightarrow0$ limit, $v_B^-$ and $v_B^+$ does not coincide since this limit does not necessarily correspond to $\frac{B}{2A}\rightarrow0$. This is because we are working with both charged and rotating black holes. Although we consider the $a=0$ case, $v_B^-$ and $v_B^+$ also do not coincide again due to the existence of the electromagnetic charges. Only when $a,q,p,\mathcal{L}\rightarrow0$, the butterfly velocities $v_B^\pm$ differ solely by a negative sign.
\section{Case for the Ultraspinning Black Holes}\label{sec5}
In this section, we would like to calculate the butterfly velocity of the ultraspinning counterparts of both Kerr-NUT-AdS and Kerr-Sen-AdS black holes. The ultraspinning limit is obtained by first redefining the coordinate $\varphi\rightarrow\varphi/\Xi$ and then taking the limit $a\rightarrow l$ \cite{Wu2021}. Following previous derivation of the Kruskal-like metric (see also \cite{Prihadi2023}), the solution for the ultraspinning black hole at $\theta=\pi/2$ is given by
\begin{align}\label{metricUVhat}
d\hat{s}^2=&\frac{\hat{F}}{\hat{\kappa}^2\hat{U}\hat{V}}d\hat{U}d\hat{V}\\\nonumber
&+\hat{h}(r)\bigg(d\hat{z}+\frac{\hat{h}_{\hat{\tau}}(r)}{2\hat{\kappa}\hat{U}\hat{V}}(\hat{U}d\hat{V}-\hat{V}d\hat{U})\bigg)^2.
\end{align}
The hatted functions are all given by \cite{Prihadi2023}
\begin{align}
\hat{F}(r)&=\frac{\hat{\Delta}(r^2+\Upsilon^2)}{\hat{\tilde{f}}^2},\\
\hat{h}(r)&=\hat{\eta}^2\frac{\hat{\tilde{f}}^2}{r^2+\Upsilon^2},\\
\hat{h}_{\hat{\tau}}(r)&=\hat{\eta}^{-1}\bigg[\frac{-\hat{\Delta}(\mathcal{L}-l)+l(\mathcal{L}l-(r^2+\Upsilon^2+l^2))}{\hat{\tilde{f}}^2}+\hat{\gamma}\bigg],
\end{align}
with
\begin{align}
\hat{\tilde{f}}^2&=-\hat{\Delta}(\mathcal{L}-l)^2+(\mathcal{L}l-(r^2+\Upsilon^2+l^2))^2,\\\nonumber
\hat{\eta}&=\frac{1}{1-\hat{\Omega}_\varphi\mathcal{L}},\;\;\;\;\;\hat{\gamma}=\frac{\hat{\Omega_\varphi}}{1-\hat{\Omega}_\varphi\mathcal{L}}.
\end{align}
\indent The location of the asymptotic for the ultraspinning case is also modified and it is not symply obtained by taking the limit $a\rightarrow l$ of the radius $\bar{r}_c(\mathcal{L})$ in eq. (\ref{radasymptotics}). The asymptotic values of the hatted functions obtained by taking $r\rightarrow r_c\gg1$ are given by
\begin{align}
\hat{\tilde{f}}(r_c)^2&\approx\frac{r_c^4\mathcal{L}}{l}\bigg(2-\frac{\mathcal{L}}{l}\bigg),\\\nonumber
\hat{h}(r_c)&\approx\frac{r_c^2\frac{\mathcal{L}}{l}\big(2-\frac{\mathcal{L}}{l}\big)}{(1-\hat{\Omega}_\varphi\mathcal{L})^2},\nonumber\\
\hat{h}_\tau(r_c)&\approx(1-\hat{\Omega}_\varphi\mathcal{L})\bigg[-\frac{l^2}{\mathcal{L}}\frac{(1-\mathcal{L}/l)}{(2-\mathcal{L}/l)}+\frac{\hat{\Omega}_\varphi}{1-\hat{\Omega}_\varphi\mathcal{L}}\bigg],\\\nonumber
\hat{F}(r_c)&\approx\frac{r_c^2l/\mathcal{L}}{(2-\mathcal{L}/l)}.
\end{align}
From here, the metric of the ultraspinning Kerr-AdS black hole in the asymptotic analogous to eq. (\ref{asympmetric}) is given by
\begin{align}
d\hat{s}^2=&\frac{r_c^2l/\mathcal{L}}{(2-\mathcal{L}/l)}\bigg[d\hat{\tau}^2\\\nonumber
&+\bigg(\frac{\mathcal{L}/l(2-\mathcal{L}/l)}{(1-\hat{\Omega}_\varphi\mathcal{L})}\bigg)^2(d\hat{z}-\hat{\omega}d\hat{\tau})^2\bigg].
\end{align}
Therefore, for the ultraspinning case, the asymptotic radius $\hat{r}_c(\mathcal{L})$ is given by
\begin{equation}
\hat{r}_c(\mathcal{L})=\frac{\mathcal{L}/l(2-\mathcal{L}/l)}{(1-\hat{\Omega}_\varphi\mathcal{L})}.
\end{equation}
The result $\hat{r}_c(\mathcal{L})$ differs from the asymptotic radius of the standard Kerr-AdS black hole $\bar{r}_c(\mathcal{L})$ in eq. (\ref{radasymptotics}). The asymptotic radius $\hat{r}_c(\mathcal{L})$ is not simply obtained by taking the $a\rightarrow l$ limit of $\bar{r}_c(\mathcal{L})$. Furthermore, the radius $\hat{r}_c(\mathcal{L})$ vanishes when $\mathcal{L}\rightarrow0$, while the radius $\bar{r}_c(\mathcal{L})$ survives the limit. This result will determine the behavior of the butterfly velocity for the ultraspinning black holes.\\
\indent The equation of motion for the shift function $\hat{f}(z)$ in the ultraspinning case also follows from the standard Kerr-AdS black holes, but with all functions replaced by the hatted functions. This is because the metric of the ultraspinning black holes in the Kruskal coordinates is in the same form with the standard Kerr-AdS counterpart (see eq. (\ref{metricUV}) for $\theta=\pi/2$ and eq. (\ref{metricUVhat})). Therefore, the butterfly velocity for the ultraspinning case is given by
\begin{equation}
\hat{v}_{B}^{\pm}=\frac{\mathcal{L}/l(2-\mathcal{L}/l)}{(1-\hat{\Omega}_\varphi\mathcal{L})}\frac{\hat{\kappa}}{|\hat{w}_{\pm}|},
\end{equation}
where now $\hat{w}_\pm$ is given by
\begin{equation}
\hat{w}_{\pm}=\frac{\hat{B}}{2\hat{A}}\pm\sqrt{\bigg(\frac{\hat{B}}{2\hat{A}}\bigg)^2+\frac{\hat{C}}{\hat{A}}},
\end{equation}
with
\begin{align}
\hat{A}=\frac{1}{2\hat{\kappa}^2\hat{h}(r_+)},\;\;\;\hat{B}=\frac{\hat{h}'_{\hat{\tau}}(r_+)}{\hat{\kappa} \hat{F}'(r_+)},\;\;\;\hat{C}=\frac{\hat{h}'(r_+)}{2\hat{h}(r_+)}\frac{1}{\hat{F}'(r_+)}.
\end{align}
\begin{figure}
\centering
\includegraphics[scale=0.7]{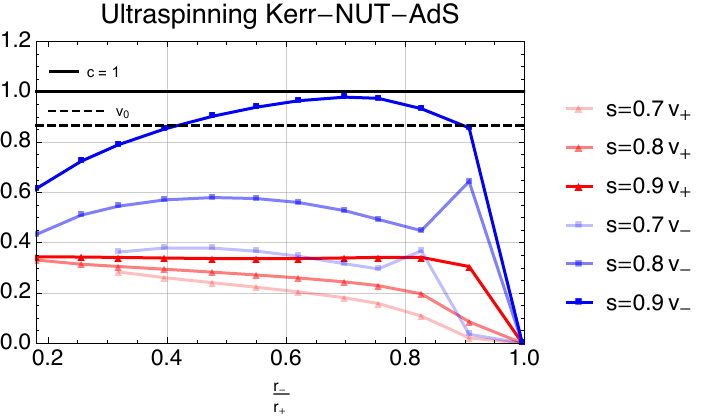}
\includegraphics[scale=0.7]{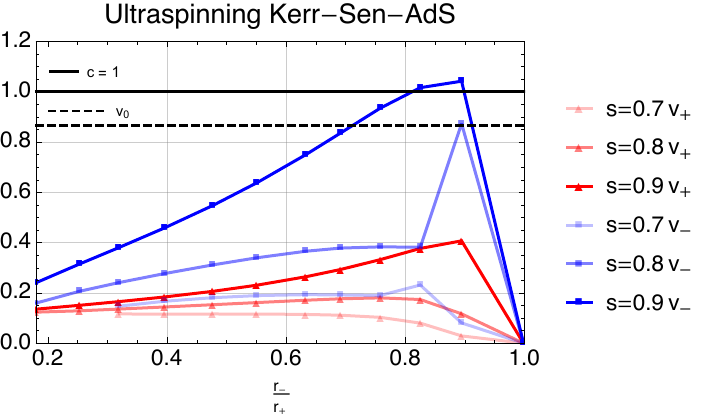}
\caption{\label{fig:ButterflyUspinKerrSen}Plot for the butterfly velocities $\hat{v}_B^\pm$ of the ultraspinning Kerr-NUT-AdS (top)  with $n=0.05$ and Kerr-Sen-AdS (bottom) black hole. The angular momentum $\mathcal{L}$ is again scaled as $\mathcal{L}=s\mathfrak{\tau}/\hat{\Omega}_\varphi$. We plot for $s=\{0.7, 0.8, 0.9\}$ from $s=0.7$ which is depicted by the line with lowest opacity up to $s=0.9$ which is the one with solid line. Blue lines represent $\hat{v}_B^-$ while red lines represent $\hat{v}_B^+$. In this case we use $p=0.2,q=0.1, l=1$.}
\end{figure}
\indent The butterfly velocities $\hat{v}_B^\pm$ for the ultraspinning Kerr-AdS black hole is shown in Figure \ref{fig:ButterflyUspinKerrSen}. It can be seen that the negative modes $\hat{v}_B^-$ are larger compared to the positive modes $\hat{v}_B^+$ although in this case, we do not observe any velocities that violate the speed of light. The butterfly velocities for the ultraspinning case also approach zero at extremality, similar to the standard Kerr-AdS black hole cases. As expected from the $\hat{r}_c(\mathcal{L})$ dependence on $\mathcal{L}$, the butterfly velocities are larger with larger value of $\mathcal{L}$. 
\section{Lyapunov Exponent}\label{sec6}
It has been shown recently \cite{Prihadi2023} that a more general rotating black hole in AdS with extra conserved charges also exhibits chaos. Using the holographic calculation of the mutual information, the scrambling time can be obtained to have a logarithmic nature $\tau_*\sim\log S$ at the leading term or large entropy limit with $r_+/l\gg1$. Furthermore, the minimum instantaneous Lyapunov exponent \cite{Malvimat2023KerrAdS4} can also be obtained from
\begin{equation}
\lambda_L=\frac{4\pi\sqrt{-F(r_\star)h(r_\star)}}{\mathcal{A}_H},
\end{equation}
where $r_\star$ is the solution to $\frac{d}{dr}(F(r)h(r))\big|_{r=r_\star}=0$ and $\mathcal{A}_H$ is the area of the horizon.\\
\indent In this section, we show how the minimum instantaneous Lyapunov exponent varies with respect to the ratio $\mathfrak{r}\equiv r_-/r_+$ ranging from 0 to 1 (extreme limit). We also plot the surface gravity analog $\kappa$ and see whether the bound $\lambda_L\leq\kappa$ is still obeyed. We choose the black hole parameters such that the large entropy limit $r_+/l\gg1$ can still be used. The result is shown in Figure \ref{LyapunovNUT} for the Kerr-NUT-AdS black hole and Figure \ref{LyapunovSen} for the Kerr-Sen-AdS counterpart. In contrast with the Lyapunov plot in \cite{Prihadi2023}, in this work, we make sure that the value of $r_+/l$ is always greater than one, although it becomes closer to one as we approach the extreme limit $\mathfrak{r}\rightarrow1$. \\
\indent It can be shown from the graph that the behavior of the Lyapunov exponent $\lambda_L$ between the Kerr-Sen-AdS and Kerr-NUT-AdS black holes are quite different, especially for large values of $\mathcal{L}$. The value of $\mathcal{C}\equiv\kappa/\lambda_L$ are $\mathcal{C}=7.7363$ for the Kerr-NUT-AdS black hole with $\mathfrak{r}=0.582711$ and $\mathcal{C}=1.53271$ for the Kerr-Sen-AdS black hole with $\mathfrak{r}=0.55617$. We may conclude that the Kerr-Sen-AdS black hole is more chaotic than the Kerr-NUT-AdS black hole.\\
\indent In the extremal limits, only the ones with $s=1$ approach some finite non-zero values of $\lambda_L$ while the other ones go to zero. This fact, along with the observation that the butterfly velocity also approaches zero in the extremal limit, shows that entanglement disruption cannot spread and will not scramble the TFD state in the extremal limit. 
\begin{center}
\begin{figure}
\includegraphics[scale=0.53]{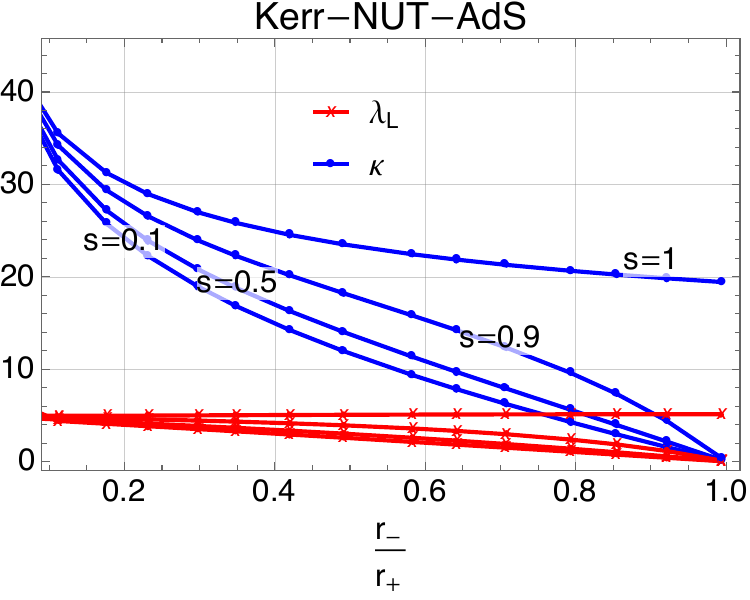}
\caption{\label{LyapunovNUT}Plot of both $\kappa$ and $\lambda_L$ with $\mathcal{L}=s\mathfrak{r}/\mu$ for the Kerr-NUT-AdS black hole. The parameters of the black hole are $a=0.05,l=0.1,p=0.2,q=0.1, n=0.03$. The ratio $r_+/l>1$ is also maintained. The range of the mass parameter is $m\in[1.5,0.3994]$ and the range of the ratio $r_+/l$ is $r_+/l\in[2.84701,1.02866]$.}
\end{figure}
\begin{figure}
\includegraphics[scale=0.53]{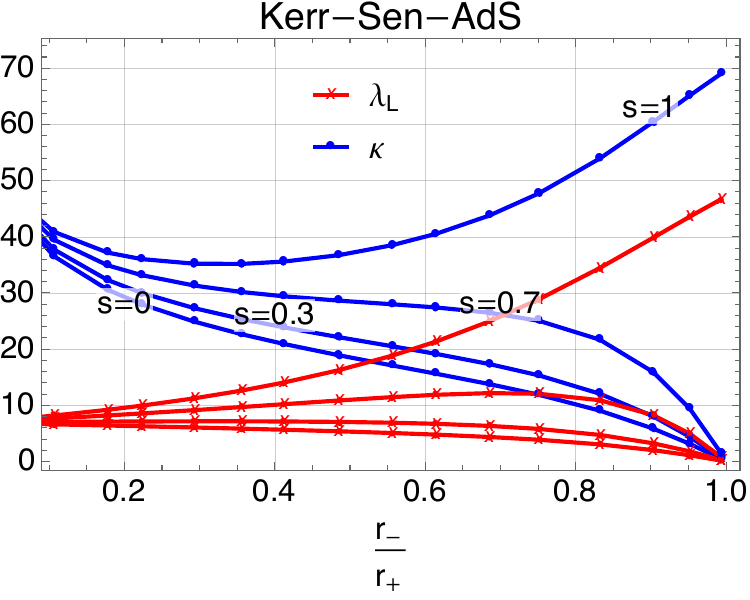}
\caption{\label{LyapunovSen}Similar plot of both $\kappa$ and $\lambda_L$ with $\mathcal{L}=s\mathfrak{r}/\mu$ for the Kerr-Sen-AdS black hole. We choose $a=0.05,l=0.1,q=0.2,p=0.1$. We then vary the mass from $m=1.5$ where $\mathfrak{r}\rightarrow0$ to $m=0.24905$ where we approach extremality. The ratio $r_+/l>1$ is always maintained, ranging from $r_+/l=2.91344$ to $r_+/l=1.19766$.}
\end{figure}
\end{center}
\vspace{-1cm}
\section{Conclusions and Discussions}\label{sec7}
The rotating and charged Kerr-AdS black hole families are chaotic under the perturbations with localized rotating and charged shock waves. The localized perturbations disrupt the entanglement pattern encoded in the TFD state and make the OTOC vanish inside the effective ”light cone” created by the scrambling time and the butterfly velocity. We show that aside from the logarithmic behavior of the scrambling time which indicates fast scrambling, the butterfly velocity can also be obtained using holography. It is shown that there are two modes of the butterfly velocity. One of the butterfly velocities obeys the causality bound $c=1$ as well as the Schwarzschild-AdS limit $v_{B0}=\sqrt{3}/2$ and the other one can exceed the speed of light and thus violate causality.\\
\indent This result might indicate that only one mode of butterfly velocity is physical. This is unique to the rotating charged black hole case since the extra factor $\frac{B}{2A}$ breaks the symmetry between $v_B^-$ and $v_B^+$. This phenomenon was also observed in \cite{Jahnke2019} for the three-dimensional BTZ black hole and it turns out that the four-dimensional Kerr-AdS black hole families are no exception. We may suspect that the superluminal behavior of one of the butterfly velocity modes also appears in other rotating AdS black holes such as the ones with higher curvature terms of even higher dimensional rotating black holes. This will be investigated in future works.\\
\indent The holographic calculation of the butterfly velocity for the rotating charged Kerr-AdS black hole families in four dimensions is important to further study the properties of the dual CFT. By knowing the black hole calculation first, it may open valuable insights to the microscopic description of the dual three-dimensional CFT which is described earlier by the entangled TFD state with extra chemical potentials due to rotation and charges. The angular momentum $\mathcal{L}$ and charges $\mathcal{Q},\mathcal{P}$ also play important roles in the chaotic properties of the black hole and hence also provide more insights to the microscopic CFT dual description.\\
\indent In this work, we show that the angular momentum of the shock waves $\mathcal{L}$ increases both the butterfly velocity and the Lyapunov exponent. We also find that the butterfly velocity approaches zero in the extremal limit, along with the Lyapunov exponent. From here, we can conclude that scrambling does not happen in the extremal limit of Kerr-AdS black hole families. This result will give us more insight into understanding chaos for extremal rotating and charged black holes and their dual theories.\\
\indent The electric and magnetic charges of the shock waves $\mathcal{Q},\mathcal{P}$ defined in \cite{Prihadi2023} do not have a direct influence on both butterfly velocities and the Lyapunov exponent (and even the scrambling time $\tau_*$). However, the charges play a vital role in delaying the scrambling process of the black hole \cite{Horowitz2022}. The charge of the shock waves interacts with the black hole charge, creating a "bounce" inside the black hole horizon so that the shock waves change their direction in the interior. The difference between the time of the bounce $\tau_d$ and the time when the shock waves are sent from the boundary $\tau_0$ denotes the scrambling delay time and the shock wave charges highly influence it. For the four-dimensional Kerr-AdS black hole families, the result is given by Eq. (109) of \cite{Prihadi2023}, which also applies to other various Kerr-AdS black hole families.\\
\indent It is also interesting to study chaos for black holes other than AdS. However, this needs a different approach than the TFD state description of the entangled dual CFT. For instance, as mentioned earlier, rotating black holes have some CFT description near extremality near its horizon via the Kerr/CFT \cite{Guica2009, Sakti2018,Sakti2019, Sakti2020b,
Sakti2020,Sakti2021,Sakti2022}. Investigating chaos due to small perturbations in this framework may also be interesting, which is left for future works.
\begin{acknowledgments}
F. P. Z. would like to thank Kemenristek, the Ministry of Research, Technology, and Higher Education, Republic of Indonesia for financial support. H. L. P. would like to thank Ganesha Talent Assistantship Institut Teknologi Bandung for financial support. H. L. P. would like to thank the members of the Theoretical Physics Groups of Institut Teknologi Bandung for their hospitality. D. D. is supported financially by Directorate of of Talent Management BRIN (Badan Riset dan Inovasi Nasional) for his postdoctoral fellowship. S. A. was supported by an appointment to the Young Scientist Training Program at the Asia Pacific Center for Theoretical Physics (APCTP) through the Science and Technology Promotion Fund and Lottery Fund of the Korean Government. This was also supported by the Korean Local Governments - Gyeongsangbuk-do Province and Pohang City.
\end{acknowledgments}
\bibliography{EEBHSRT.bib}
\appendix
\newpage

\end{document}